\documentclass[dvipdfmx]{pasj01}
\usepackage{graphicx}
\draft

\begin{document}

\title{What Determines Unique Spectra of Super-Eddington Accretors?: Origin of Optically Thick and Low Temperature Coronae in Super-Eddington Accretion Flows}
\author{Norita \textsc{Kawanaka}\altaffilmark{1,2} and Shin \textsc{Mineshige}\altaffilmark{1}}
\altaffiltext{1}{Department of Astronomy, Kyoto University, Kitashirakawa-Oiwake-cho, \\
Sakyo-ku, Kyoto 606-8502, Japan}
\altaffiltext{2}{Hakubi Center, Kyoto University, Yoshida-Honmachi, \\
Sakyo-ku, Kyoto 606-8501, Japan}
\email{norita@kusastro.kyoto-u.ac.jp}

\maketitle
\begin{abstract}
Existence of relatively cool ($k_B T \lesssim 10~{\rm keV}$) and optically thick ($\tau \gtrsim 3$) coronae are inferred above super-Eddington accretion flow such as ultraluminous X-ray sources (ULXs), GRS 1915+105, and narrow-line Seyfert 1 galaxies (NLS1), which contrasts the cases in sub-Eddington accretion flows, which are associated with coronae with $k_B T \sim 100~{\rm keV}$ and $\tau \sim 1$.  To understand their physical origin, we investigate the emission properties of the corona which is formed by the gas blown off the super-Eddington inner disk by radiation pressure.  We assume that the corona is heated by the reconnection of magnetic loops emerged from the underlying disk.  We show that this radiation pressure driven wind can act as an optically thick corona which upscatters thermal soft photons from the underlying disk, and that with a reasonable parameter set we can theoretically reproduce the coronal optical depth and temperature which are inferred by spectral fittings of observational data.  By contrast, the coronal optical depth cannot be so high in sub-Eddington cases, since the coronal material is supplied from the disk via evaporation and there is a maximum limit on the evaporation rate.  We support that the low temperature, optically thick Comptonization should be a key signature of super-Eddington accretion flow.
\end{abstract}

\section{Introduction}

Recently, the existence of super-Eddington (or supercritical) accretion flows onto black holes have been investigated from both observational and theoretical points of view (see, e.g., \cite{kato+08}).  For example, ultraluminous X-ray sources (ULXs) are very bright X-ray sources whose X-ray luminosity is much larger than the Eddington luminosity of a stellar-mass black holes $L_X\gtrsim 10^{39}{\rm erg}~{\rm s}^{-1}$, and two possibilities have been presented to account for ULXs: sub-Eddington accretion onto an intermediate mass black hole (IMBHs; \cite{colbertmushotzky99, makishima+00, miller+03, miller+04, cropper+04}) and super-Eddington accretion onto a stellar-mass black hole (\cite{watarai+01, king+01, poutanen+07}).  In fact, some studies show that their spectral features can be well explained by the model with super-Eddington accretion flows (\cite{ebisawa+03, okajima+06, tsunoda+06, vierdayanti+06, vierdayanti+08, gladstone+09, kawashima+12, 2013MNRAS.435.1758S, 2013Natur.503..500L}).   In addition, Narrow-Line Seyfert 1 galaxies (NLS1s) are considered to harbor a black hole with a super-Eddington accretion flow (e.g. \cite{zhouwang99, mineshige+00, collinkawaguchi04, 2017MNRAS.471..706J}).  As for Galactic X-ray sources, some microquasars (e.g. GRS 1915+105; \cite{done+04, done+07, vierdayanti+10b}) are the candidates of super-Eddington accreting black holes.  The detailed observational data of these sources would help us interpret comprehensively the behavior of super-Eddington accretion flows. 

Despite extensive spectral analyses of potential super-Eddington sources the presence of super-Eddington accretion flow is still a controversial issue, since there is no good theory to predict its firm observational signatures.  We definitely need information of black hole masses.  One exceptional case is GRS 1915+105, for which the black hole mass is known to be $14\pm 4 M_{\odot}$ \citep{greiner+01}.  This is the best target in our Galaxy to study high luminosity states of accretion \citep{done+07}.  The other exceptional cases are ULX pulsars, whose central objects are accreting neutron stars and thus less massive than $3M_{\odot}$.  To explain their luminosities ($\gtrsim 10^{40}~{\rm erg}~{\rm s}^{-1}$) super-Eddington accretion is required, and eight ULX pulsars have been discovered so far \citep{2014Natur.514..202B, 2016ApJ...831L..14F, 2017MNRAS.466L..48I, 2017Sci...355..817I, 2018MNRAS.476L..45C, 2018ApJ...863....9W, 2020ApJ...895...60R, 2019MNRAS.488L..35S, 2020MNRAS.495.2664C}. 

As for the theoretical approach, simple analytical or semi-analytical modeling have been attempted under one-zone approximation (e.g. \cite{shakurasunyaev73, abramowicz+88, honma+91, poutanen+07}).  To study the gas dynamics of super-Eddington accretion flow, global, multi-dimensional, radiation-hydrodynamic (RHD) simulations are indispensable.  Such simulations have been attempted by several authors \citep{eggum+88, okuda02, okuda+05, ohsuga+05}.  Especially \citet{ohsuga+05} performed large simulation box, long-term RHD simulations of super-Eddington accretion flows for the first time, and have demonstrated that super-Eddington accretion onto a black hole in a quasi-steady fashion is feasible.  Moreover, they have shown that such super-Eddington disks naturally give rise to high-velocity ($v \sim 0.1-0.2c$) outflows driven by radiation pressure\footnote{Note that the possibility of such super-Eddington outflows was already suggested by \citet{shakurasunyaev73}.}.  Importantly, there is no upper limit to the mass accretion rate to a black hole nor to the mass outflow rate; the mass accretion rate and outflow rate are roughly in proportion to the mass supply rate \citep{ohsuga07}.  Note, however, these studies adopted the $\alpha$-prescription for the viscosity in the accretion flows.  This shortcoming was improved by \citet{ohsuga+09}, who have presented for the first time global two-dimensional radiative magnetohydrodynamic simulations of black hole accretion flows following the angular momentum transports via the magnetic torque, and these simulations have also shown the existence of the high-velocity disk outflows driven by radiation-pressure force in super-Eddington accretion flows (see also \cite{ohsugamineshige11}).

The outflow associated with a super-Eddington accretion flow may be optically-thick with respect of Thomson scattering for sufficiently high $\dot{M}>L_{\rm Edd}/c^2$, where $L_{\rm Edd}$ is the Eddington luminosity, and electrons in it would Compton-upscatter soft photons from the underlying accretion flow.  The plasma surrounding an accretion disk and upscattering soft photons from the disk is called a corona and is considered to exist universally in accretion disks associated with AGNs and X-ray binaries that are emitting hard X-rays \citep{1976SvAL....2..191B, 1977A&A....59..111B, sunyaevtruemper79}, and its temperature is supposed to be $k_{\rm B}T \sim 100~{\rm keV}$ as a result of energy dissipation of magnetic fields amplified in the accretion disk \citep{haardtmaraschi91}.  \citet{kawashima+09} showed by the RHD simulation that the Compton $y$ parameter of the radiation pressure-driven outflow from a super-Eddington accretion flow is around unity.  That is, occurrence of significant Comptonization was expected.  Later, \citet{kawashima+12} calculated the radiation spectra of super-Eddington accretion flows using a Monte Carlo method and showed that the radiation pressure-driven outflow acts as a corona above the disk and should Compton upscatters soft photons from the underlying disk, thereby modifying the X-ray spectra from super-Eddington accretion flows (see \cite{kitaki+17} for a more recent work).  On the other hand, \citet{gladstone+09} did the observational studies about ULXs and found that most of their X-ray spectra can be fitted well with the thermal component being Compton upscattered by relatively low-temperature ($k_{\rm B} T \lesssim 10{\rm keV}$) and optically-thick ($\tau_{\rm sc} \gtrsim 3$) plasma.  Similar spectral fitting is also performed for other ULXs \citep{2013PASJ...65...48Y, 2017ApJ...839...46S} and for GRS 1915+105 \citep{vierdayanti+10b}, whose mass accretion rate is considered to be super- or near-Eddington as stated above, and for NLS1s \citep{2018arXiv180511314I}. Therefore, these facts support the supercritical accretion flow model for ULXs.  However, the physical interpretation of those low temperature and large optically depth of that corona has not been described so far.

In this study, we investigate the basic properties of outflowing coronae in supercritical accretion flows with a simple, semi-analytical model.  Especially we aim at understanding the physical reason for the low temperature and large optical depth of coronae associated with some ULXs, microquasars, and NLS1s.  It is obvious that the scattering optical depth can be higher for larger mass outflow rates, though it remains a question how and why electron temperatures should be low.  In the present study we raise the following questions; why electron temperatures tend to be low in the outflowing coronae from supercritical accretion flows; why $\tau$ of the outflowing corona cannot be so high in subcritical flows.  We discuss these issues from the viewpoint of energy balance in the outflowing corona.  In our calculation we take into account the heating due to magnetic reconnections and cooling due to inverse Compton scatterings in radiation pressure-driven outflows in a simple and self-consistent manner.  

\section{Model}
In the present paper, we apply the basic concept of the magnetic reconnection heated corona model by \citet{haardtmaraschi91} to the case of super-Eddington accretors.  The big distinction lies in that disk coronae are formed not via evaporation of disk gas but by radiation-pressure driven outflow.
\subsection{Coronal Density}
We consider a super-Eddington accretor with $\dot{M}> L_{\rm Edd}/(\eta c^2)$, where $\eta \sim 0.1$ is the radiation efficiency.  Emergence of a strong disk wind is expected from the innermost region of the supercritical disk, being driven by radiation-pressure force by the disk \citep{eggum+88}.  Due to this mass loss process, the mass accretion rate in the disk is effectively suppressed, and generally depends on the radius.  
%According to the standard disk model \citep{shakurasunyaev73}, the local radiation flux from the disk surface is proportional to $r^{-3}$, while the local Eddington flux $F_{\rm Edd}(r)=L_{\rm Edd}/(4\pi r^2)$ is proportional to the gravitational force $\propto r^{-2}$.  Therefore, we can assume the radial dependence of the mass accretion rate as
It is often assumed that the local accretion rate in an outflowing accretion disk varies as a power law in radius,
\begin{eqnarray}
\dot{M}(r)=
\left\{ \begin{array}{ll}
\dot{M}_0 & r>r_{\rm crit} \\
\dot{M}_0 \left(\frac{r}{r_{\rm crit}} \right)^s & r\le r_{\rm crit}. \label{mdot}
\end{array} \right.
\end{eqnarray}
Here $s$ is a constant less than unity and the critical radius $r_{\rm crit}$ is determined as \citep{ohsuga+05}
\begin{eqnarray}
r_{\rm crit}=\frac{3\dot{M}_0 c^2}{4L_{\rm Edd}}r_{\rm S},
\end{eqnarray}
where $r_{\rm S}=2GM_{\rm BH}/c^2$ is the Schwarzschild radius, and $\dot{M}_0$ is the mass accretion rate at the outer edge.  Assuming that the wind velocity is on the same order of the escape velocity $v_{\rm esc}=(2GM_{\rm BH}/r)^{1/2}$, the coronal density distribution $n_{\rm cor}(r)$ should satisfy
\begin{eqnarray}
m_{\rm p} n_{\rm cor} v_{\rm esc} 4\pi r dr=d\dot{M}(r),
\end{eqnarray}
which gives
\begin{eqnarray}
n_{\rm cor}(r)=\frac{s}{4\pi m_{\rm p} rv_{\rm esc}}\frac{\dot{M}(r)}{r},
\end{eqnarray}
where $m_{\rm p}$ is the proton mass.  The scattering optical depth of an outflowing corona can be obtained as
\begin{eqnarray}
\tau_{\rm cor}(r)=n_{\rm cor}(r)\sigma_T \ell_{\rm cor}, \label{opticaldepth}
\end{eqnarray}
where $\ell_{\rm cor}$ is the scale height of the outflowing corona, which will be given in the following section.

\subsection{Optical Depth and Temperature of a Corona}
In this subsection, we present the heating process for the outflowing corona above a supercritical accretion disk.  In the context of a subcritical accretion flow, \citet{haardtmaraschi91} proposed a corona model in which the corona is heated by the reconnection of magnetic loops emerging from the rotating disk, and here we discuss the energy balance of an outflowing corona following their idea.  As for the strength of magnetic field, here we assume that its energy density in the disk is the constant fraction of the radiation energy density in the disk.  Then the magnetic field in the disk can be determined by
\begin{eqnarray}
\frac{B^2}{8\pi}\approx \eta_B\cdot {\rm max}\left( aT_{\rm disk}^4, \frac{\dot{M}\Omega}{2\pi r \alpha}\right),
\end{eqnarray}
where $\eta_B<1$ is a parameter which should be given from the results of radiative magnetohydrodynamic simulations (e.g. \cite{ohsugamineshige11}) in the following discussion.  Here $aT_{\rm eff}^4$ and $\frac{\dot{M}\Omega}{2\pi r \alpha}$ are the radiation pressure and gas pressure (\cite{2013ApJ...766...31K}).  Here $T_{\rm disk}$ is the temperature at the equatorial plane, which is given from the standard model of a radiation pressure-dominated disk \citep{shakurasunyaev73} as
\begin{eqnarray}
T_{\rm disk}=2.1 \times 10^7{\rm K} \left(\alpha_{-1} M_{{\rm BH},1}\right) ^{-1/4} r_{1}^{-3/8}, \label{tdisk}
\end{eqnarray}
where $M_{{\rm BH},1}=M_{\rm BH}/10M_{\odot}$, $\alpha=10^{-1}\alpha_{-1}$ is the viscous paramter, and $r=10r_{\rm S} r_{1}$ is the radial distance from a black hole.  We note that this temperature is independent of the mass accretion rate.  Then we can estimate the magnetic field strength in the case with a radiation-pressure dominated disk:
\begin{eqnarray}
B\approx 1.9 \times 10^7~{\rm Gauss}~\eta_{B,-2}^{1/2} \alpha_{-1}^{-1/2} M_{{\rm BH},1}^{-1/2} r_{1}^{-3/4},
\end{eqnarray}
where $\eta_{B,-2}=\eta_B/10^{-2}$.

Following \citet{liu+02}, we assume that the energy equilibrium of the outflowing corona is established between the magnetic reconnection heating and the inverse Compton cooling.  The vertical scale height of an outflowing corona is determined by the loop length of the magnetic field emerged from the underlying disk, $\ell_{\rm loop}$.  Here we assume that $\ell_{\rm loop}$ has the same order as the radial position of the spot where the magnetic loop emerged: $\ell_{\rm loop}\simeq r$.  In the innermost region, however, the loop should remain long enough due to the existence of a funnel structure in the outflow \citep{kitaki+17}, and so we set the floor value of the loop length as $\ell_{\rm loop}={\rm max}(r,10r_{\rm S})$.  In addition, we should take into account the fact that the corona above the super-Eddington flow is not static but is upwardly moving.   Photons can escape the outflowing corona when their diffusion timescale, $\ell_{\rm cor}\tau_{\rm cor}/c$, is shorter than the escape timescale for the corona, $\ell_{\rm cor}/v_{\rm esc}$, which gives the lower limit for the scale height of the corona as
\begin{eqnarray}
\ell_{\rm cor,lim}\approx \frac{c}{v_{\rm esc}n_{\rm cor}\sigma_T}.
\end{eqnarray}
Hereafter we estimate the coronal scale height as $\ell_{\rm cor}\approx {\rm min}(\ell_{\rm cor, lim}, \ell_{\rm loop})$.  
%In the inner region of the accretion flow, where the coronal scale height is determined as $l_{c,{\rm lim}}$, the scattering optical depth of the corona along the vertical direction.

Then the coronal heating rate per unit surface area due to magnetic reconnection can be described as
\begin{eqnarray}
Q^+=\frac{B^2}{4\pi}V_R\cdot \frac{\ell_{\rm cor}}{\ell_{\rm loop}}, \label{coronalheating}
\end{eqnarray}
where $V_R$ is the speed of energy dissipation via magnetic reconnection.  Here we simply assume that $V_R={\rm min}(V_A,c)$ where $V_A=B/(4\pi n_{\rm cor} m_{\rm p})^{1/2}$ is the Alfv\'{e}n velocity.  On the other hand, the coronal cooling is dominated by inverse Compton scattering of seed photon field supplied from the underlying disk:
\begin{eqnarray}
Q^-=\frac{4k_B(T_{\rm cor}-T_{\rm rad})}{m_e c^2}cU_{\rm rad}\cdot {\rm max}(\tau_{\rm cor},\tau_{\rm cor}^2), \label{coronalcooling}
\end{eqnarray}
where $U_{\rm rad}$ is the energy density of the soft photon field to be Comptonized in the corona \citep{1979rpa..book.....R}.  Here we take into consideration that the temperature of coronal electrons $T_{\rm cor}$ may not be much higher than that of soft photon field, $T_{\rm rad}=(U_{\rm rad}/a)^{1/4}$, and that we cannot neglect the term $4k_B T_{\rm rad}$ in the numerator of the right-hand side.  Since the radiation energy density from a super-Eddington accretion flow is limited by the local Eddington flux, we can assume the soft photon field emitted intrinsically from the underlying disk as 
\begin{eqnarray}
U_{\rm rad, intr}&=&\frac{L_{\rm Edd}}{4\pi r^2 c}\simeq 3.7\times 10^{12}~{\rm erg}~{\rm cm}^{-3}M_{{\rm BH},1}^{-1}r_{1}^{-2}. \label{trad}
\end{eqnarray}
As the soft photon field, we should also take into account the reprocessed radiation, $U_{\rm rad, re}$, that originates from the backward Compton scattering in the corona,
\begin{eqnarray}
U_{\rm rad, re}=F_{\rm cor, irr}/c, \label{reprocess}
\end{eqnarray}
where $F_{\rm c, irr}$ is the irradiation flux from the corona onto the underlying disk.  Here the disk albedo is assumed to be unity.  The soft photon field is reevaluated as $U_{\rm rad}=U_{\rm rad, intr}+U_{\rm rad, re}$, which should have the Planck distribution.  By solving $Q^+=Q^-$ at each radius, we can evaluate the radial distribution of coronal temperature.  We should note that $F_{\rm c,irr}$ can be obtained only after solving the radiation transfer inside the corona by Monte Carlo simulations (see Sec.\ref{sec:MC}) for a calculated temporal temperature distribution of the corona.   Therefore, to obtain the coronal temperature distribution consistently, we should repeat performing Monte Carlo simulations and updating the coronal temperature until it converges.

Let us estimate the typical optical depth of the wind corona in our model.  Here we adopt $s=0.15$, which is implied by recent radiative hydrodynamic simulations of super-Eddington accretion flows \citet{kitaki+18}.  The coronal density as a function of radius, black hole mass, and mass accretion rate can be then evaluated as
\begin{eqnarray}
n_{\rm cor}(r)\simeq 1.2\times10^{17}~{\rm cm}^{-3}r_{1}^{-1.35}M_{{\rm BH},1}^{-1}\dot{M}_{1}^{0.85},
\end{eqnarray}
where $\dot{M}_{1}=\dot{M}_0/(10\dot{M}_{\rm Edd})$, and $\dot{M}_{\rm Edd}=L_{\rm Edd}/\eta c^2$ is the Eddington accretion rate, where $\eta=0.1$ is the radiation efficiency.  The optical depth can be then described as
\begin{eqnarray}
\tau_{\rm cor}\simeq 2.4r_{1}^{-0.35}\dot{M}_{1}^{0.85}\left( \frac{\ell_{\rm cor}}{r} \right),
\end{eqnarray}
which can be larger than unity when the mass accretion rate is well above the Eddington value.  This is consistent with the observations of ULXs and other super-Eddington accreting black holes.

\section{Results}
\subsection{Coronal Structure} \label{sec:MC}
In our model, soft thermal photons are emitted from an optically-thick disk, and some of them are Compton upscattered by hot electrons in the outflowing corona lying above the disk.  As a result, the radiation spectrum emerged from such a system would have a hard component in addition to the thermal component originating from a disk.

We use Monte Carlo simulations to calculate the radiation spectrum expected from our model of a wind-fed disk corona, based on the prescription presented in \citet{pozdnyakov+77}.   As mentioned in the previous subsection, the coronal temperature distribution and radiation spectrum are obtained iteratively.  First, we assume at each radius the seed photon field, $U_{\rm rad}$ as blackbody radiation whose energy density is given by Eq.(\ref{trad}).   Second, we compute the coronal temperature and optical depth at that radius as described in the previous section.  Third, we solve the radiation transfer in a corona by Monte Carlo simulations.  Here we adopt the vertically one-zone approximation for the corona at each radius.  Especially, we take into account the bulk motion of a corona, whose upward velocity is given by the escape velocity at that radius.  Fourth,  we obtain the downward flux from the corona (i.e., the irradiation flux onto the underlying disk, $F_{\rm cor, irr}$), and evaluate the reprocessed photon field by Eq.(\ref{reprocess}).  Fifth, we compare the reprocessed photon field, $U_{\rm rad,re}$, and the photon field used in the simulations, $U_{\rm rad}$.  If the former is comparable to or larger than the latter, we should reevaluate the soft photon field as $U_{\rm rad}=U_{\rm rad, intr}+U_{\rm rad, re}$, and go back to the second step to perform the Monte Carlo simulations with renewed coronal temperature.  If the former is much smaller than the latter at each radius, we regard that the consistent solution is obtained, and compute the radiation luminosity by integrating the energy spectra of photons emerged from the upper boundary of coronae for all radii.

Fig. 1 depicts the temperature of the outflowing corona as a function of radius.  We can see that the coronal temperatures is roughly several tens of ${\rm keV}$, and we see a tendency that the higher the mass accretion rate is, the cooler becomes the coronal temperature.  Note that the outflow (corona) solution exists only in the inner regions, and that the higher the accretion rate is, the larger becomes the cutoff radius.

We see several breaks in the slope of each curve (see the arrows with numbers attached to the curve of $\dot{M}=10\dot{M}_{\rm Edd}$).  In the innermost region (i.e., inside the point (1) for $\dot{M}=10\dot{M}_{\rm Edd}$), the coronal scale height is determined by the diffusion length of a photon within the dynamical timescale, $\ell_{\rm cor,lim}$ (see Eq. 9).  At the break point (1), $\ell_{\rm cor,lim}$ is equal to the loop length, $\ell_{\rm loop}=10r_{\rm S}$, and the coronal scale height is determined as $\ell_{\rm cor}=\ell_{\rm loop}=10r_{\rm S}$ in the region between the points (1) and (2).  The magnetic loop length is switched from $10r_{\rm S}$ to $r$ at the break point (2), and the coronal scale height is determined as $\ell_{\rm cor}=\ell_{\rm loop}=r$ in the region between the points (2) and (3).  Finally, the break point (3) corresponds to the radius where the radiation pressure is equal to the gas pressure: outside this point the radiation pressure would dominate over the gas pressure.

Fig. 2 depicts the scattering optical depth of the outflowing corona as a function of radius.  In the innermost region, the coronal optical depth increases with radius as $\propto r^{1/2}$.  This dependence can be explained in the following way: in the innermost region, as shown above, the scale height of the outflowing corona is determined as $\ell_{\rm cor,lim}$ in Eq. 9.  Therefore, the optical depth is evaluated as $n_{\rm cor}\sigma_T \ell_{\rm cor,lim}\approx c/v_{\rm esc}\propto r^{1/2}$.  Note that this value is independent of the mass accretion rate, which can also be seen in Fig. 1.  The optical depth peaks at the point (2) where the loop length $\ell_{\rm loop}$ is equal to $\ell_{\rm cor,lim}$.  In the region between (1) and (2), except the case of $\dot{M}=50\dot{M}_{\rm Edd}$, $\tau_{\rm cor}$ decreases with radius steeply, $\propto r^{s-3/2}$, because the coronal scale height is evaluated as $\ell_{\rm cor}=\ell_{\rm loop}=10r_{\rm S}$ in this region.  In the outermost region, since $\ell_{\rm cor}=r$, $\tau_{\rm cor}$ decreases with radius shallowly, $\propto r^{s-1/2}$.  When the mass accretion rate is large enough, $\ell_{\rm cor,lim}$ is always smaller than the floor value of the scale height, $10r_{\rm S}$, the middle region disappears.  In contrast with the case of sub-Eddinton accretion, where the coronal optical depth is around unity, the peak value of the optical depth of outflowing corona is larger than unity, $1\lesssim \tau_{\rm cor} \lesssim 10$.  

Fig. 3 depicts the Compton $y$-parameter of the outflowing corona as a function of radius.  We can see that in the inner region where the outflowing corona is sufficiently optically-thick, the $y$-parameter is the order of unity, which is similar to the case of sub-Eddington accretion flows.

%Fig. 1, 2, and 3 depict the temperature optical depth, and Compton-$y$ parameter, $y=4k_B(T_{\rm cor}-T_{\rm rad})/m_ec^2\cdot {\rm max}(\tau_c, \tau_c^2)$, of the corona as functions of radius, respectively, predicted from our model with $\alpha=0.1$ and $\eta_B=0.015$.  We can see that the coronal temperature is a few tens of ${\rm keV}$, and that the typical optical depth of corona is $1<\tau_{\rm cor}\lesssim 10$.  Compared to sub-Eddington accretors such as typical X-ray binaries and AGNs, whose coronal temperature and optical depth are $\sim 100~{\rm keV}$ and $\sim 1$. respectively, the outflowing corona is cool and optically thick, which nicely reproduces the coronal properties of super-Eddington accretors inferred from observations of ULXs, NLS1s, and GRS 1915+105.  We can also see that the coronal temperature gets lower for higher accretion rate.

We should note that the energy conservation is not violated in our model.  Fig. 4 depicts the fraction of disk energy, $3GM_{\rm BH}\dot{M}/(8\pi r^3)$, which is dissipated in the corona via magnetic reconnection for various accretion rates, assuming $\eta_B=7.5\times 10^{-3}$.  We can see that the ratio never exceeds unity at any accretion rate and radius, which means that the energy conservation is not violated in the calculations.  Note that this fraction should be around unity throughout the corona in the case of sub-Eddinton accretion \citep{haardtmaraschi91, liu+02}.

\begin{figure*}
\begin{center}
\includegraphics[width=10cm,clip]{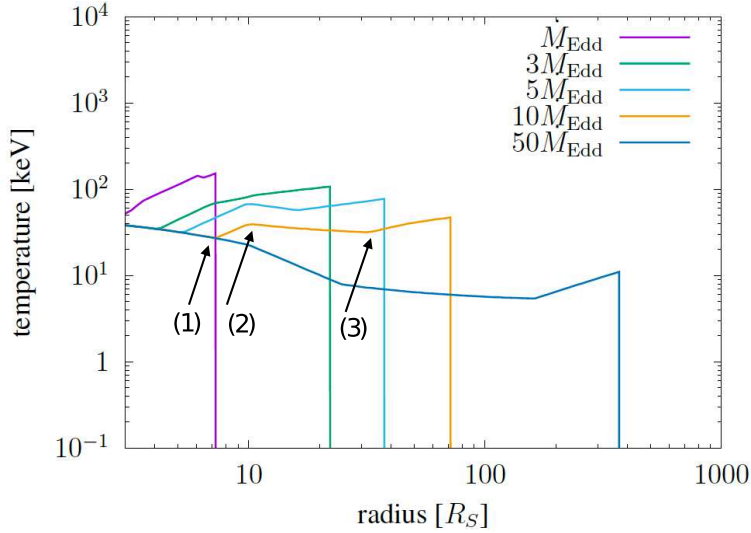}
\end{center}
\caption{Coronal temperatures as functions of radius for different accretion rates.  The black hole mass is fixed as $M_{\rm BH}=10M_{\odot}$.  The breaks in the plot of $\dot{M}=10\dot{M}_{\rm Edd}$ are indicated with arrows: the breaks (1), (2), and (3) correspond to the radii where the loop length $\ell_{\rm loop}=10r_{\rm S}$ is equal to $\ell_{\rm cor,lim}$, where the loop length $\ell_{\rm loop}=10r_{\rm S}$ is equal to $r$ (i.e., $r=10r_{\rm S}$), and where the gas pressure $\dot{M}\Omega/(2\pi r \alpha)$ is equal to the radiation pressure $aT_{\rm disk}^4$ in the disk, respectively.}
\end{figure*}

\begin{figure*}
\begin{center}
\includegraphics[width=10cm,clip]{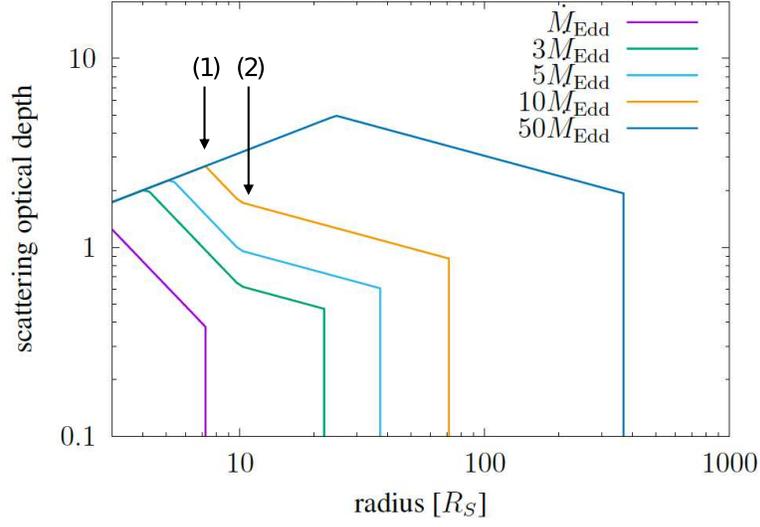}
\end{center}
\caption{Scattering optical depths of the corona as functions of radius for different accretion rates.  The black hole mass is fixed as $M_{\rm BH}=10M_{\odot}$.  The explanations of the breaks (1) and (2) in the plot of $\dot{M}=10\dot{M}_{\rm Edd}$ are the same as those in Fig. 1.}
\end{figure*}

\begin{figure*}
\begin{center}
\includegraphics[width=10cm,clip]{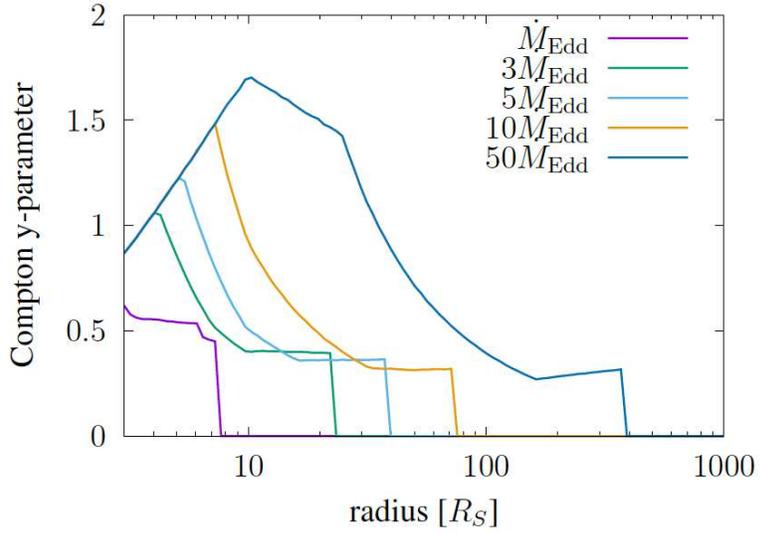}
\end{center}
\caption{Compton-$y$ parameters of the corona as functions of radius for different accretion rates.  The black hole mass is fixed as $M_{\rm BH}=10M_{\odot}$.}
\end{figure*}

\begin{figure*}
\begin{center}
\includegraphics[width=10cm,clip]{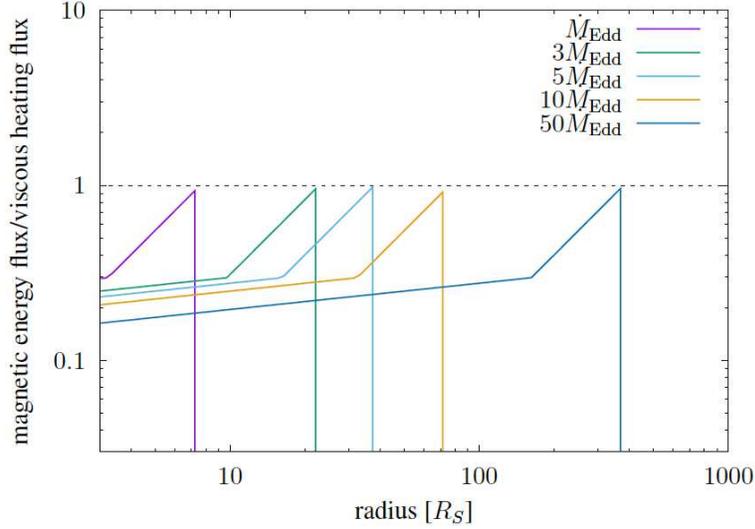}
\end{center}
\caption{The fraction of disk energy which is dissipated in the corona via magnetic reconnection as functions of radius for various accretion rates.  All the ratios do not exceed unity.}
\end{figure*}

\subsection{Radiation Spectra}
Figure 5 depicts the results of the Monte Carlo simulations for the radiation spectra expected from our corona model for different mass accretion rates.  First, we can see that each of the computed spectra consists of a soft thermal component whose peak frequency is $\sim 10^{17.4}~{\rm Hz}$ ($\sim 1~{\rm keV}$), and a hard Comptonized component whose frequency ranges up to $\sim 10^{18.5-19.5}~{\rm Hz}$ ($\sim 10-100~{\rm keV}$).  We can also see that the peak energy and cutoff energy of the Comptonized component get lower with higher accretion rate.  When the mass accretion rate is $\gtrsim 5\dot{M}_{\rm Edd}$, especially, the Comptonized component merges with the soft thermal component, and the spectrum has a broadened peak.  Such a feature is consistent with what is seen in the so-called broadened disk state \citep{2017ARA&A..55..303K}.

\begin{figure*}
\begin{center}
\includegraphics[width=10cm,clip]{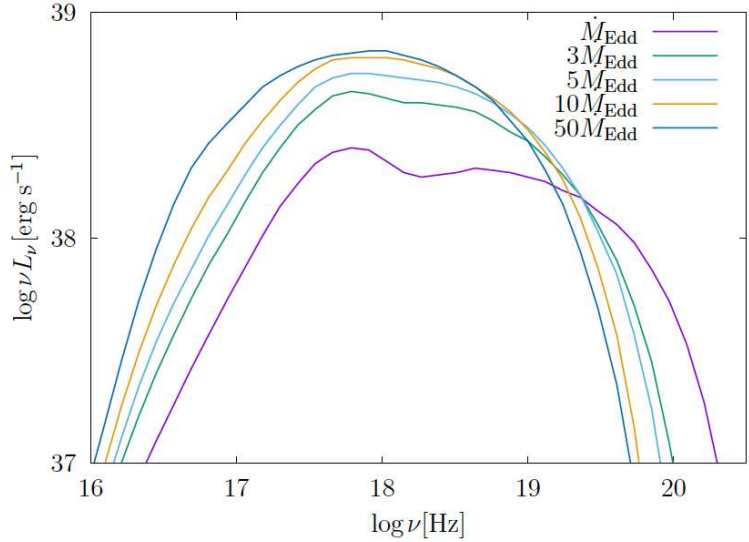}
\end{center}
\caption{Radiation spectra from super-Eddington accretion flows with wind-fed coronae for different mass accretion rates.  The parameters are $(M_{\rm BH},\alpha, \eta_B)=(10M_{\odot}, 0.1, 0.015)$}
\end{figure*}

\section{Discussion}
\subsection{Origin of an optically-thick corona}
According to the spectral analyses for BH objects suspected to harbor supercritical accretion flows done in the past studies \citep{gladstone+09, vierdayanti+10a, 2017ApJ...839...46S, 2018arXiv180511314I}, their X-ray spectra can be fitted by a thermal component and a hard X-ray component that are supposed to come from an optically-thick accretion disk and a hot corona surrounding the disk, respectively.  However, the corona assumed in these studies is cool ($k_B T\lesssim 10~{\rm keV}$) and optically thick ($\tau_e \gtrsim 3$) compared to the corona assumed in fitting the spectra of BH objects with subcritical accretion flows.

Although the formation process of corona is still unclear even in the case with subcritical accretion flows, it is often considered that the corona is build up by the evaporation of the underlying disk due to the vertical heat conduction from a hot corona to a cool disk \citep{1994A&A...288..175M, 2000A&A...361..175M, 2000MNRAS.316..473R, 2002ApJ...575..117L, liu+02, 2007MNRAS.376..435M}.  In the model proposed by \cite{liu+02}, they discuss the structure of a corona that is heated by the reconnection of magnetic fields emerged from the underlying disk.  By analogy with the solar corona \citep{2001ApJ...549.1160Y}, they consider the chromospheric evaporation at the interface of a cool disk and a corona.  When the conduction flux from a corona heats up some of the chromospheric plasma into the magnetic tube, the density of a corona is determined by the energy balance at the interface, described as the equation of
\begin{eqnarray}
\frac{k_0T_c^{7/2}}{l}\approx \frac{\gamma}{\gamma-1}nk_BT_c\left( \frac{k_B T_c}{m_H} \right)^{1/2},
\end{eqnarray}
where $k_0\approx 10^{-6}~{\rm erg}~{\rm cm}^{-1}~{\rm s}^{-1}~{\rm K}^{-7/2}$, $\gamma=5/3$, and $l$ is the length of the magnetic tube.  Combining this equation with the energy balance in the corona (i.e., $Q^+$ in Eq.\ref{coronalheating} is equal to $Q^-$ in Eq.\ref{coronalcooling}), one can derive the density and temperature of a corona as functions of radius, black hole mass, and mass accretion rate.  Especially, the scattering optical depth of a corona derived in their model is limited below unity (see their Fig. 1) on the contrary to our outflowing corona model.  This can be explained in the following way.

In the evaporation scenario, the plasma supply rate to the corona would be higher when the coronal temperature is higher.  However, when the corona becomes too dense and too optically thick with respect to electron scattering, due to the inverse Compton cooling, its temperature cannot be too high, which would regulate the evaporation rate at the interface between the corona and disk.  In the outflowing corona model, by contrast, the coronal plasma is supplied from the disk wind, whose efficiency is mainly determined by the radiation pressure inside the disk and the effect of the thermal conduction is subdominant.  Therefore, the scattering optical depth of the corona in our model is not limited to below unity as in the evaporation scenario.

Since the escape velocity in the innermost region is up to $\sim $ a few ten \% of the speed of light, the outflowing corona there is also assumed to have relativistic speed.  In such a case, the scattering optical depth of the outflowing corona $\tau_{\rm cor}$ should be modified by a factor of $\gamma (1-\beta \cos \theta)$, where $\beta=v_{\rm esc}/c$, $\gamma=(1-\beta^2)^{-1/2}$, and $\theta$ is the angle between the velocity of the outflowing corona and the line of sight \citep{1991ApJ...369..175A}.  However, in the realistic case, the outflow is accelerated gradually and reaches the relativistic speed only when it goes up to reach the height, $z\sim r$, which is on the order of the scale height of the corona.  Therefore, the modification to the coronal optical depth would be rather subtle, and so our main conclusions would not be altered.

 To fully check the energy balance one should evaluate the radiation flux and kinetic energy flux of the outflowing corona originating from the locally liberated energy in the underlying disk per unit surface area.  However, since photons generated inside the accretion flow would be significantly trapped and advected inward in the case with super-Eddington accretion, it is not straightforward to check if the results are consistent energetically only from the distributions of radiation flux.  In our calculations, moreover, the radiation flux from the underlying disk is dominated by reprocesses photons, whose energy originally comes from the magnetic energy dissipated to heat the corona.  As for the outflow, the amount of ejected material is much smaller than that of accreted material because we assume $\dot{M}\propto r^s$ where $s=0.15$, which is inferred from the latest radiation hydrodynamical simulations\citep{kitaki+18}.  Therefore, the kinetic energy flux of the outflow is negligible compared to the energy liberated in the disk.
 
\subsection{Radiation spectra}
In each radiation spectrum shown in Figure 5, the hard component (i.e., Compton upscattered component) is less energetic than the thermal component.  Such spectra are seen in some ULXs (e.g., NGC 5204 X-1, NGC 4559 X-1, NGC 5408 X-1 etc.; \cite{gladstone+09}).  However, some ULXs (e.g., IC 342 X-1; \cite{2017ApJ...839...46S}) and other BHs (e.g., GRS 1915+105; \cite{vierdayanti+10b}) with supercritical accretion flows show radiation spectra that are dominated by a hard component, and such spectra cannot be reproduced in our model as it is, because the corona is not so hot in the inner region, where most of seed photons are emerged from the disk.  However, if we take into account the radial motion of the outflowing corona, we would have different results.   Actually, the outflowing corona would suffer from the centrifugal force and be expected to move radially outward.  In such a case, the whole corona would be more optically thick and as a result, the corona would be cooled down to lower temperature via efficient inverse Compton scattering.

Another effect that may produce a hard component in the spectrum is the formation of the funnel wall.   The radiation-hydrodynamic simulations of supercritical accretion flows done in \citet{kitaki+17} show that the collision between the inflowing material and the funnel wall (i.e., the low-density region near the polar axis) produces overheated regions whose temperature is up to $\sim 10^8~{\rm K}$ in the vicinity of a black hole, and that significant amount of hard photons are produced there.  This is beyond the scope of the simple analysis made in this paper.
 
\section{Summary}

We construct a simple model of an outflowing corona formed in a black hole accretion flow whose mass accretion rate is comparable to or above the Eddington value.  In this model, the coronal plasma is fed by the radiation pressure-driven wind from an underlying disk, and heated by the reconnection of magnetic loops emerged from the disk.  By solving the energy equation of this wind-fed corona taking into account the cooling due to the Compton scattering of soft photons emitted from a disk, we found that the coronal temperature is relatively low ($k_B T_c \sim 10~{\rm keV}$), and that the coronal optical depth is relatively large ($\tau_c \gtrsim 3$) compared to the corona expected in subcritical accretion flows, which is independent of the black hole mass.  The existence of such a cool and optically-thick corona has been inferred from the spectral fitting of GRS 1915+105, some ULXs, and some NLS1s.  We calculate the radiation spectra emerged from this corona by Monte Carlo simulations, and show that the spectral features of these sources are qualitatively reproduced in our corona model.  Since such an optically-thick corona is not likely to be formed in a subcritical accretion flow, where the coronal plasma is fed by the chromospheric evaporation driven by the heat conduction from a hot corona to a cool disk, because when the corona becomes dense and optically-thick, it would be efficiently cooled via inverse Compton scattering, which would suppress the chromospheric evaporation.  Since the corona in our model is fed by the disk wind that is driven by the radiation pressure in the disk, it can avoid such a regulation process and can be sufficiently optically-thick.  Our results support that the cool and optically-thick coronae whose existence is inferred from the X-ray spectra observed in some ULXs, microquasars, and NLS1 are the key signature of supercritical accretion flows.

We appreciate the anonymous referee for his/her helpful comments and suggestions.  This work is supported in part by JSPS KAKENHI Grant Number, 20K04026 (S. M.).  N.K. is supported by the Hakubi project at Kyoto University.

\end{document}